\documentstyle[preprint,aps,epsfig]{revtex}
\begin{document}
\newcommand{\beq}{\begin{equation}}
\newcommand{\eeq}{\end{equation}}
\narrowtext
\begin{center}
\LARGE
Nondiagonal Parton Distributions in the Leading Logarithmic Approximation

\bigskip
\large
L.L.Frankfurt$^a$, A.Freund$^b$, V.Guzey$^b$, M. Strikman$^b$

\normalsize
$^a$Physics Department, Tel-Aviv University, Tel-Aviv, Israel\\ 
$^b$Department of Physics, Penn State University\\
University Park, PA  16802, U.S.A.
\end{center}
\vskip .3in
\begin{abstract}
In this paper we make predictions for nondiagonal parton distributions in
a proton in the LLA.  We calculate the DGLAP-type evolution kernels in 
the LLA, solve the nondiagonal GLAP evolution equations with a modified 
version of the CTEQ-package and comment on the range of applicability 
of the LLA in the asymmetric regime. We show that the nondiagonal gluon 
distribution $g(x_{1},x_{2},t,\mu^2)$ can be well approximated at small $x$ 
by the conventional gluon density $xG(x,\mu^2)$. 
\end{abstract}
   
\section{Introduction}

Due to the experimental possibility of probing nondiagonal distributions 
in hard diffractive electro-production processes, theoretical interest 
in this area in recent years 
\cite{Brod'94,FKS'95,Rad'96,C.F.S'96,Ji'96,Radpriv,Abram'95} has produced 
interesting results.
A pioneering analysis of the nondiagonal distributions for  
the diffractive photoproduction of $Z^0$-bosons in DIS where the 
applicability of PQCD is guaranteed was given by Bartels and Loewe in 1982 
\cite{Bartels} but went essentially unnoticed.

In the this paper we would like 
to complement these results by concrete predictions, albeit to the LLA, 
which can be tested by an experiment.  In Sec.\ \ref{sec:asym}
we shall demonstrate that in the limit of  small $x$ the amplitudes of
hard  diffractive processes can be calculated in terms of discontinuities 
of nondiagonal parton distributions. The real part of the amplitude will be 
calculated by applying a dispersion representation of the amplitude over
$x$. We will show that the term in the amplitude which cannot be calculated in 
terms of the discontinuities of nondiagonal parton distributions 
\cite{Rad'96,C.F.S'96} is suppressed by one power of $x$ in this
limit. This result will make it possible to calculate the evolution 
kernels in the LLA following the traditional methods \cite{Dok'80} and to 
compare them to results obtained in the QCD-string operator 
approach \cite{B.B'88}.  

In Sec.\ \ref{sec:kernel} we calculate the nondiagonal kernels and find them 
equivalent to those in \cite{Rad'96,Ji'96}. They are different from the 
evolution equations for nondiagonal parton densities which were presented
without derivation in \cite{Levin}. 
In Sec.\ \ref{sec:pred} we shall make predictions about the 
nondiagonal parton distributions by solving, numerically, the nondiagonal GLAP
evolution equations with the help of a modified version of the 
CTEQ-package. In Sec.\ \ref{sec:lim} we shall discuss the limitations of the 
approximation and the need for NLO-results. Future directions will be
discussed in the conclusions.

\section{Nondiagonal parton distributions and hard diffractive processes.}
\label{sec:asym}

It has been recently understood that the major difference in QCD
between leading twist effects in DIS and higher-twist effects in 
hard diffractive processes is to be attributed to the fact that the 
latter, initiated by highly virtual, longitudinally polarized
photons, can be calculated in terms of nondiagonal, rather than diagonal, 
parton distributions \cite{Abram'95}.

Thus, in order to calculate unambiguously hard two-body processes, 
it is necessary to calculate nondiagonal parton distributions in a 
nucleon. This implies knowledge of the non-perturbative nondiagonal parton 
distributions in the nucleon which have not been measured so far. Hence, the
aim of this section is to express the nondiagonal parton distributions in the 
nucleon through quantities being maximally close to the diagonal
parton distributions. Our second aim is to elucidate on the kinematics of the 
nondiagonal parton distributions in the nucleon needed to describe hard 
diffractive processes. We shall also discuss the expected, limiting, 
behaviour of the nondiagonal parton distributions. 

For the leading twist effects QCD evolution equations have traditionally 
been discussed in terms of the imaginary part of the amplitude. 
This is because the bulk of experimental data available is on the total 
cross sections of inclusive processes. However it is well known that 
the QCD evolution equation has a simple form for the whole amplitude 
which includes both real and imaginary parts \cite{GribLip}.  
This form of the evolution equation can be generalized to the case of 
higher-twist processes, hard diffractive processes \cite{FKS'95} and 
hard two-body processes\cite{C.F.S'96}. The analysis of the QCD evolution 
equation for the nondiagonal parton densities shows that the evolution equation
contains two terms. The first one is described by a DGLAP-type evolution 
equation\cite{FKS'95,Rad'96,C.F.S'96,Abram'95}, whereas the second term,
found in Ref.\ \cite{Rad'96} for vector meson production at 
small $x$, cannot be interpreted in terms of parton distributions. The QCD 
evolution of this term is governed by the Brodsky-Lepage evolution equation
\cite{Rad'96,Radpriv}.

\subsection{GLAP evolution equation for hard diffractive processes}
 
The aim of this section is to prove that for hard diffractive processes
in general, the $Q^2$-evolution
at any $x$ in the DGLAP-region as discussed below, is described by a 
nondiagonal GLAP-type evolution equation with asymmetric DGLAP-type kernels 
and that these processes can be 
calculated through the discontinuity of hard amplitudes. This property is 
important for the quantitative
calculations since the dispersive contribution has a relatively simple
physical interpretation and a deep relation with the conventional
parton densities. As to the first step, we shall deduce a relationship 
between amplitudes of hard two-body processes and parton densities,
and we will find an additional term which has no probabilistic interpretation.
We will restrict ourselves to the $Q^2$-region where the parton distributions 
are still rising and the additional term is of no importance as discussed 
below.

The QCD factorization theorem for hard processes means that the hard
blob can be factorized from the soft one with a precision of a power
of $Q^2$. The topologically dominant Feynman diagrams for small $x$ processes
correspond to attachments of only two gluons to the hard blob.
Although our analysis is rather general, for certainty we shall
restrict ourselves to the case of diffractive processes where diagrams 
with two-gluon exchange dominate.\footnote{ 
Hard collisions due to the exchange of 2 quarks are numerically small in the 
LLA at small $x$.}
It is convenient to decompose the 
momentum of the exchanged gluon $k$ in Sudakov-type variables: 
\beq
k=x_{1} \tilde p  +\beta \tilde q  +k_{t},
\eeq
where 
\beq
\tilde p^2=\tilde q^2=0\ \ \mbox{and}\ \ (k_{t} q)=(k_{t} p)=0.
\eeq 
To express the amplitude in terms of non-diagonal parton distributions,
the contour of integration over $\beta$ should be closed over the 
singularities of the amplitude in gluon-nucleon scattering at fixed $x_1$ 
and $x_1-x$. The singularities over $\beta$ are located in the complex plane
of  discontinuities over the gluon virtualities: 
${i\epsilon\over x_1}$ and ${i\epsilon\over x_1-x}$, and from the
$s$- and $u$- channel discontinuities: ${-i\epsilon\over(1-x_1)}$ and
${i\epsilon\over(1+x_1-x)}$. 
The amplitude differs from 0 if these singularities pinch the contour 
of integration. This causality condition restricts the region of 
integration to: 
\beq 
-1+x\leq x_{1} \leq 1.
\eeq
Our main interest is in the amplitude in the physical region where 
$-t\geq 0$ but small as compared to other relevant scales of the process
under consideration. In this region the amplitude can be represented 
as the sum of terms having $s$- or $u$-channel singularities only.
For the $s$-channel contribution to the amplitude of hard 
diffractive processes,
given $1\geq x_{1} \geq x$, the integral over $\beta$  can only be closed over
the discontinuity in the amplitude of the gluon-nucleon scattering 
in the variable $s$. 

Thus this contribution to the amplitude is expressed through the
imaginary part  of the amplitude for gluon-nucleon scattering.
The QCD evolution of this term is described by a GLAP-type 
evolution equation where the kernel accounts for the off-diagonal kinematics. 
One also has to add a similar term corresponding to $u$-channel singularities.

The contribution of the region $x \geq x_1 \geq 0$ has no direct 
relationship to the conventional parton densities. This is 
because the integral over $\beta$  cannot be closed for $s$- or $u$-channel 
discontinuities but it may be closed for the  discontinuities over the 
gluon "mass". In Ref.\ \cite{Rad'96} the analogy of this term with the wave
function of a vector meson has been suggested. The presence of this
piece which cannot be evaluated in terms of parton densities
introduces theoretical uncertainties into the treatment of hard 
two-body processes at large and moderate $x$.

However, for the imaginary part of the amplitude, more severe 
restrictions on the region of integration apply: 
\beq
1 \geq x_{1} \geq x.
\eeq  
In order to ensure that we have not included any superfluous regions of 
integration, we consider the discontinuity of the hard amplitude in the $s$-
channel. This additional restriction follows from the
requirement that the mass of a $q\bar q$ system in the intermediate 
state should be larger than the mass of a  produced vector meson such that 
the hadron cannot decay into the $q\bar q$ system.
This result helps to prove that the piece which cannot be evaluated in terms
of parton densities is
inessential for hard diffractive processes. Let us now
apply a dispersion representation over the variable $s$.  The only term
which cannot be reconstructed in terms of a dispersion relation,
i.e in terms of discontinuities of parton densities,
is the subtraction constant\footnote{This constant is independent of $s$.} 
in the real part. The
contribution of the subtraction term to the amplitude with a positive 
signature, i.e symmetric under the transposition of  $s\rightarrow u$, 
is suppressed by an additional power of $s$ or, equivalently, by an additional
power of $x$. For the processes with negative charge parity in the 
crossed channel, i.e.  electroproduction of a neutral pion, the amplitude is 
antisymmetric under the transposition $s\rightarrow u$\footnote{This 
corresponds to a negative signature.}.
This amplitude has no subtraction terms at all, since, in QCD, it increases 
with energy slower than s\footnote{An odderon-type contribution in PQCD
is suppressed by an additional power of $Q^2$.}. Thus, in this case, 
a dispersion
representation gives the full description. To summarize let us point out once
more that the small $x$ behaviour of hard diffractive processes is described 
through the discontinuities of hard amplitudes.

\subsection{Small $x_{i}$ behaviour of the nondiagonal gluon distribution}

We want to stress that the slope of the $x$ dependence of the amplitudes for 
diffractive processes, however not their residue, should be independent of
the asymmetry between fractions $x_1$ and $x_2$ of the nucleon momentum carried
 by the initial and final gluons. This is due to the fact that
the  $x_i$ of the partons in the ladder are essential , but not the external 
$x$, and increase with the length of the parton ladder. Hence, the 
asymmetry between the 
gluons may be important in one or two rungs of the ladder but not 
in the whole ladder. Therefore, at sufficiently small $x$, it is legitimate to 
neglect $x$ in most of the rungs of the ladder as compared to the 
$x_{i}$. This means that the asymmetry between the gluons influences the 
residue but not the slope of the $x$ dependence.

Let us now discuss the small $x_{i}$ behaviour of 
$g(x_{1},x_{2})$ -- the nondiagonal gluon density in a nucleon.
The factorization theorem -- Eq. 3 of  \cite{C.F.S'96}-- is the basis for 
the formal definition of the nondiagonal gluon density as the matrix
element of gauge-invariant bilocal operators (cf. eq. 6 of \cite{C.F.S'96}):
\begin{eqnarray}
   g_{g/p}(x_{1},x_{2},t,\mu )  &=&
   - \int _{-\infty }^{\infty } \frac {dy^{-}}{4\pi }
   \, \frac {1}{p_{+}^2}
   \;
   e^{-ix_{2}p^{+}y^{-}}
   \langle p'|\; T G_{\nu }{}^{+}(0,y^{-},{\bf 0}_{T}) \,
     {\cal P} \, G^{\nu +}(0)\; |p\rangle .
\label{eq.gluondis} 
\end{eqnarray}

Here $\cal P$ is a path ordered exponential of a gluon field along 
the light-like
line joining the two gluon operators, $t$ is the square of 
invariant momentum transferred to the target, and $\mu$ describes the scale 
dependence. The sum over transverse gluon polarizations is implied. 
Actually Eq.\ \ref{eq.gluondis} coincides with the definition given in 
\cite{Rad'96,C.F.S'96} for the same quantity. 

For $x_{1}=x_{2}$, $g_{g/p}(x_{1},x_{2})$ is related to the diagonal gluon 
distribution as $xG_{{\rm diag}}(x)=g_{{\rm nondiag}}(x,x)$. Within the 
leading $\alpha_s \ln x$ approximation
where the difference between  $\ln x_{i}$  and $\ln x$ 
can be neglected, this distribution coincides with the diagonal one 
\cite{Brod'94}:  
\beq 
G_{{\rm leading\, \alpha_s \ln x}}(x_{1},x_{2},t,\mu^2)= xG(x,\mu^2).
\eeq 

We want to stress here that at fixed Bjorken variable  $x$, the cross 
sections of 
hard diffractive processes are expressed through $g(x_{1},x_{2},t,\mu^2)$  
where $x_1-x_2=x$. This can be proved by calculating the
high energy limit of hard diffractive processes and then applying Ward
identities similar to Ref.\ \cite{Brod'94}. This means that the region of 
integration near $x_2=0$  ($x_2\ll x_1$) gives only a
small contribution to the amplitudes of hard diffractive processes.

Alternatively, one can examine the leading regions of the integrals 
in the calculation of the distribution of a parton in a parton 
which is imperative 
in finding the correct hard scattering coefficients for the desired process.
This calculation is necessary since one has not only ultraviolet divergences 
in the partonic cross sections from which one wants to extract the 
Wilson coefficients, but also infrared divergences stemming from initial-state 
collinearities of the participating partons (see Ref.\ \cite{Muell'89}
for further details) which are cancelled by the perturbatively calculated 
expansion of the parton distribution. The region of 
$x_{2}=0$ does not give a leading contribution. This can be seen by 
using a simple argument that proper Feynman diagrams have no singularity at 
$x_{2}=0$, and the region of integration over the exchanged gluon momenta
$x_{2}=0$ forms an insignificant part of the permitted phase volume.

In the first step one has to show that a gluon with $x_2=0$ corresponds 
to a soft gluon and then one can use the argument by Collins and Sterman 
\cite{ColSter} first introduced for proving factorization in inclusive 
$e^+e^-$-reactions:
\begin{itemize}

\item For clarification, the quark-loop to which the gluons attach consists
      both of the hard part and the part whose momenta are parallel to 
      the vector meson and of the order ${\nu\over m_N} = {2qp_N\over m_N}$.

\item The minus component $l_-$ of the quark-loop
      which is transferred to the target proton is
      $l_-\approx m_N x$. The minus component of the gluon momentum $k$ is:
      \beq k_- \simeq \frac{(m_{q\bar q}^2 +M_V^2)m_N}{2qp_N}
      \ll l_-,
      \eeq
      where $m_{q\bar q}$ is the mass of the $q\bar q$ system. 
      Thus we can neglect $k_-$ with respect to $l_-$ in the
      calculation of the leading term of the amplitude corresponding to 
      the leading power in the energy of the process.

\item The transverse momentum in the quark loop ($l_t$) is cut off by the 
      vector meson wave function and thus $l_t^2\ll Q^2$ in stark contrast 
      to DIS, whereas the $k_t^2$ of the gluon is only restricted by the 
      virtualities in the photon wave function which can be as high as $Q^2$
      \footnote{ The similarity to DIS will be restored at extremely large 
      $Q^2$ as a consequence of both a Sudakov-type form factor in the photon 
      vertex and a slow decrease, with increasing $k_t$, of the vector meson
      wave function.}.          
      However one has to satisfy the Ward-identity $k_{\mu}T^{\mu\nu} = 0$
      where $T$ stands for the amplitude. Using Sudakov-variables this becomes:
      \beq
      x_2p_{\mu}T^{\mu\nu} + k_{t\mu}T^{\mu\nu} = 0,
      \eeq
      thus for $x_2\rightarrow 0$ the transverse momentum of the gluon is very
      small and can be safely neglected as compared to $l_t$.

\item One concludes from the above said that the $x_2=0$ region corresponds 
      to a soft gluon ($k^2\sim 0$) and we can use the argument by 
      Collins and Sterman.   

\end{itemize}

Keeping the above said in mind, the $k$-integral involved in the determination 
of the leading regions is of the form (up to overall factors):
\begin{center}
\begin{eqnarray}
\int_{soft\, k}d^4 k\frac{1}{((l-k)^2+i\epsilon)(k^2+i\epsilon)}f(l-k,p)
\nonumber\\
\simeq \int_{soft\, k}dk_+ \frac{1}{((l_+ - k_+)l_{-}-l^2_t + i\epsilon)
(2k_+k_{-}-k^2_t + i\epsilon)}f(l-k,p),
\end{eqnarray}
\end{center}
where $f(l-k,p)$ is the amplitude of gluon-nucleon scattering and the 
integrals over $k_{-}$ and $k_{t}^2$ are suppressed for convenience.
If one now integrates over the remaining $k_+$ momentum one will have the 
following situations:
\begin{itemize}
\item $k_+k_-\geq k^2_t$: There are no obstructions in the 
      deformation of the integration contour since $l_t\gg k_t$
      and therefore the pole at $k^2=0$ is far from the region of integration
      allowed in the LLA.
      In other words this region does not give a leading contribution.
\item $k_+k_-\ll k^2_t$:  There are no obstructions to the contour 
      deformation since $l_t\gg k_t$ or in other words $k_+k_-$ can be 
      neglected as compared to $k_t^2$ since the  singularities do not pinch 
      the contour of integration. Thus this region does not give a 
      leading contribution either.
\end{itemize}

Thus one has proved that if one of the gluons attaching the soft to the hard 
part has $x_{2}=0$, it will be soft and thus, according to the above 
reasoning, the $x_2=0$ region of integration does not give a leading 
contribution to the parton distribution.

\section{Kernels in the LLA}
\label{sec:kernel}

There are several possible ways of calculating the evolution kernels to 
leading order in QCD. We first used the traditional approach of calculating 
the evolution kernels in the LLA, via the method of decay cells of e.g., 
a quark decaying into a quark \cite{Dok'92}\footnote{
Changes appropriate to the nondiagonal case were made.
} and using cut-diagram techniques to calculate the appropriate Feynman graphs.

As a cross-check we calculated the first order corrections to the bi-local 
quark and gluon operators on the light-cone, which not only yielded the 
nondiagonal kernels for the DGLAP equation but also the nondiagonal 
Brodsky-Lepage kernels since we were calculating the whole amplitude, not only
its imaginary part. However, since we are not interested in those kernels at 
the moment we will not comment on this fact further, let it be said though 
that our results on the Brodsky-Lepage kernels agree with those of
 Ref.\ \cite{Ji'96,Radpriv}.

We performed the calculation of the cross-check in a planar gauge 
i.e $q' \cdot A = 0$ with $q'^2\neq 0$\footnote
{ The advantage of such a physical gauge being that no gluons couple to the 
operators to first order, simplifying the calculations considerably. }
and used once more Sudakov-variables:
\beq
k = \beta p' + \alpha q'  +k_{t},
\eeq
where 
\beq
p'^2 = q'^2=0\ \ \mbox{and}\ \ (k_{t} q)=(k_{t} p)=0.
\eeq 
Since one is neglecting the proton mass one can set $p' = P$ where $P$
is the proton momentum.

The insertion of the appropriate bi-local operators for quarks and gluons 
on the light cone into the Feynman graphs for first order corrections to those
operators, short circuits the $+$-momentum in the graph, which means that the 
loop variable $k$ has not $+$-momentum $\beta P$ but rather $x_1 P$ 
(see \cite{Muell'89} for more details on calculating one loop corrections to 
parton dsitributions.). This fact eliviates us from the duty of taking the 
integral over $\beta$. In the calculation of the kernels, it remains to take 
the integral over $\alpha$ which can be done by taking the residues and then 
isolating the leading term multiplying $dk^2_{t}/k^2_{t}$ and the tree level 
amplitude. This will then yield the kernels in 
the leading logarithmic approximation.\footnote{ Note that the quark to quark 
and gluon to gluon kernels also need the self-energy diagrams to regulate 
the two possible collinear singularities, of course, after proper 
renormalization.} 

In the integral over $\alpha$, one finds three different residues. Two 
residues stemming from the vertical quark or gluon propagators which yield 
$\alpha = \frac{k^2_{t}}{x_1 s}$ and $\alpha = \frac{k^2_{t}}{x_2 s}$ giving
a contribution in the Brodsky-Lepage region, i.e the Brodsky-Lepage kernels,
 and one stemming from the horizontal propagator yielding 
$\alpha = \frac{k^2_{t}}{(y_1-x_1)s}$ which contributes to the DGLAP region,
i.e the DGLAP kernels. This is analogous to the statements made in 
Sec.\ \ref{sec:asym}.     

After having taken proper care of the definitions of our quark and gluon 
distributions in the amplitudes, we find the following expressions for the
nondiagonal evolution kernels, where $\Delta$ is given by $x_{1} - x_{2}$  
and corresponds to $x_{B_{j}}$ of e.g., vector-meson production.  
For the quark $\rightarrow$ quark transition we get:
\beq 
P_{qq}(x_1,\Delta)=\frac{\alpha_s}{\pi} 2 C_f\left [ \frac{x_1 + x_1^{3} - 
\Delta (x_1 + x_1^{2})}
{(1 - \Delta) (1 - x_1)} -\delta (1-x_1)\left [\int^{1}_{0} 
\frac{dz_1}{z_1} + \int^{1}_{0}\frac{dz_2}{z_2} -\frac{3}{2} 
\right ] \right ],
\eeq
The other kernels are computed the same way from the appropriate diagrams:
\begin{eqnarray}
P_{qg} (x_1, \Delta) &=& \frac{\alpha_s}{\pi} N_F \frac{[x_1^{3} + 
x_1(1 - x_1)^{2} - x_1^{2} \Delta]}{(1 - \Delta)^2},\\
P_{gq}(x_1, \Delta) &=& \frac{\alpha_s}{\pi} C_F \frac{[ 1 + (1-x_1)^{2} - 
\Delta]}{1 - \Delta},\\
P_{gg}(x_1, \Delta)&=&2 N_{c} [\frac{(1 - x_1)^2 + (\frac{1}{2} - x_1^{2})
(x_1 - \Delta)}{(1 - \Delta)^2} - \frac{1}{2} - \frac{x_1}{2} + 
\frac{1}{2}\frac{1}{1 - x_1} + \frac{1}{2}\frac{x_1 - \Delta}
{(1 - x_1)(1 - \Delta)}\nonumber\\  
& & +\delta (1-x_1) \left [ \frac{\beta_0}{4N_C} - \frac{1}{2}\int^{1}_{0} 
\frac{dz_1}{z_1} - \frac{1}{2}\int^{1}_{0} \frac{dz_2}{z_2} \right ]].
\label{eq.kern}
\end{eqnarray}
A word conccerning our regularization prescription is in order. In 
convoluting the 
above kernels, after appropriate scaling of $x_1$ and $\Delta$ with $y_1$,
with a nondiagonal parton density, one has to replace $z_1$ and $z_2$ in the 
regularization integrals with $z_1 \rightarrow (y_1 - x_1)/y_1$ and $z_2 
\rightarrow (y_1-\Delta)/(y_1-x_1)$. This leads to the following 
regularization prescription as employed in the modified version in the CTEQ 
package in the next section and in agreement with Ref.\ \cite{Radpriv}:
\begin{eqnarray}
\int^{1}_{x_1}\frac{dy_1}{y_1}\frac{f(y_1)}{1-x_1/y_1}_{+} &=& 
\int^{1}_{x_1}\frac{dy_1}{y_1}\frac{y_1f(y_1)-x_1f(x_1)}{y_1-x_1} 
+ f(x)\ln (1-x_1)\\
\int^{1}_{x_1}dy_1\frac{(x_1-\Delta)f(y_1)}{(y_1-x_1)(y_1-\Delta)}_{+} &=& 
\int^{1}_{x_1}\frac{dy_1}{y_1}\frac{y_1f(y_1)-x_1f(x_1)}{y_1-x_1}  - 
\int^{1}_{x_1}\frac{dy_1}{y_1}\frac{y_1f(y_1)-\Delta f(x_1)}
{y_1-\Delta}\nonumber\\
& & + f(x_1)\ln \left ( \frac{1-x_1}{1 - \Delta} \right )
\end{eqnarray}  

For $\Delta = 0$ one obtains, necessarily, the diagonal kernels, however for
the distributions $q = x_1 Q(x_1,Q^2)$ and $g = x_1 G(x_1,Q^2)$, since we
chose the definitions of our nondiagonal distributions to go into $q$ an $g$ 
rather then $Q$ and $g$. We have cross-checked these results with those of 
Ref.\ \cite{B.B'88} via the conversion formulas given by Radyushkin in a 
recent paper \cite{Rad'96}. The formulas given in a recent paper by 
Ji \cite{Ji'96} do not seem to agree with ours but this is only due to a 
different choice of independent variables used by Ji. After appropriate 
transformations, the formulas of \cite{Ji'96} agree with our results 
\cite{Radpriv}. It should be noted however that the kernels from Ref.\ 
\cite{Rad'96,Ji'96,Radpriv} are given for $Q$ and $g$ and not for $q$ and $g$ 
as we do. One just has to multiply the kernels given by those authors for the 
quark evolution equations with $x_1/y_1$ after appropriate changes for 
independent variables of course. Conversion formulas between the different 
notations can be found in Ref.\ \cite{Radpriv}.  

The evolution equation for the quantities $g(x_1,\Delta)$
and $q(x_1,\Delta)$ take the following form in our notation:
\begin{eqnarray}
& &\frac{dg(x_{1},\Delta,Q^2)}{d\ln{Q^2}}=\int^{1}_{x_{1}} 
\frac{dy_{1}}{y_{1}}\left [ P_{gg}g(y_{1},\Delta,Q_0^2)+P_{gq}q(y_{1},
\Delta,Q_0^2)\right ]
\nonumber\\
& &\frac{dq(x_{1},\Delta,Q^2)}{d\ln{Q^2}}=\int^{1}_{x_{1}} 
\frac{dy_{1}}{y_{1}}\left [ P_{qq}q(y_{1},\Delta,Q_0^2)+P_{qg}g(y_{1},
\Delta,Q_0^2)\right ].
\label{eq:evol}
\end{eqnarray} 

 We are interested in the calculation of the asymptotic
distribution in terms of the symmetric distribution in the limit 
of small $x$ and large $Q^2$. The reason why this is possible
is that in this limit the main contribution originates 
from the non-diagonal distributions at 
$\tilde x_{1},\tilde x_{2}=\tilde x_{1}-\Delta$ with 
$\tilde x_1 \gg x_1$. In the case 
$\tilde x_1, \tilde x_2 \gg \Delta$ deviations from the diagonal 
distribution are small and can be neglected.

In the following section we will present the results of our numerical
study and show that for the case of $x_{1}\gg x_{2}\simeq 0$
in the kinematic region of practical interest the diagonal and nondiagonal 
distribution will coincide for large $Q$ up to about a factor of $2$.

\section{Predictions for nondiagonal parton distributions}
\label{sec:pred}
Utilizing a modified version of the CTEQ-package, we calculate the 
evolution of the nondiagonal 
parton distributions, starting from a low $Q_{0}$=1.6 GeV and 
with rather flat initial distributions for the diagonal and 
nondiagonal case by using the most recent global CTEQ-fit CTEQ4\cite{cteq4}.
The difference between the initial-diagonal and the initial-nondiagonal 
distribution is the factor $x$ multiplying the nondiagonal distribution 
i.e $g(x_1,x_2) = x_1 G(x_1)$ in the normalization point, in accordance with 
our earlier argument that the possible difference in the distributions at 
small $x$ and large $Q$ is only given by the $Q^2$-evolution of 
$g(x_1,x_2,t,\mu^2)$.

We have only considered light quarks, since we are interested in a
proton as the initial state hadron and the $s$-quarks 
are only considered to give a small correction. 
The following figures (see Fig.\ \ref{Fig.1}) show the ratio of the 
nondiagonal distribution $g(x_1,x_2)$ to the diagonal distribution $xG(x)$
from $Q$=7 GeV to $Q$ =110 GeV and 
$x_2$ from $\frac{x_1}{100}$  
to $x_1$ with $x_1$ = $1.1\,10^{-4}$, $1.1\,10^{-3}$, $1.1\,10^{-2}$. 
 
The nondiagonal and diagonal distributions agree for 
$x_2 \rightarrow x_1$, i.e. for vanishing asymmetry, as expected, and within
a deviation of a factor between $0.2$ and $1.7$, they agree for 
$x_2\ll x_1$. The expectation that there is no $\ln x_2$, which would give a 
singluarity for $x_2 \rightarrow 0$, is also supported by our numerical 
calculations. In fact if one takes the $x_2=0$ limit, we find firstly 
(see Fig.\ \ref{Fig.2}) that the ratio of nondiagonal to diagonal distribution is finite 
i.e no $\ln x_2$ infinity and secondly that the evolution of the
nondiagonal distribution differs significantly in size and shape from the 
diagonal distribution as first anticipated by Radyushkin in \cite{Radpriv}.
   
Note that at large $Q^2$ and fixed $\Delta\ll 1$ $g(x_{1},x_{2})$ is 
determined by the initial parton distributions 
at $x_1,x_2 \gg \Delta$ where the validity of the diagonal approximation for 
$g(x_{1},x_{2})$ does not depend on our argument in Sec. \ref{sec:asym}.
The numerical calculation finds that the ratio of nondiagonal to diagonal 
distribution is larger than $1$ as 
anticipated by Radyushkin \cite{radpriv1} based on general arguments about 
the nature of the double distribution which he discusses in \cite{Radpriv}.

To see whether our numbers i.e. our numerical methods could be trusted, we 
used the MATHEMATICA program to calculate the first iteration and the first 
derivative of the evolution to see how good or bad our numbers were. 
As it turns out our integration routines produce a very good agreement with 
the numbers from MATHEMATICA with a relative difference of $5\%$. This leads 
us to believe that our numbers can be trusted to high accuracy for $x_2$ of 
$O(x_1)$ and within $5\%$ at $x_2$ down by two orders of magnitude as compared
to $x_1$.

A few words about the nature of the modifications to the CTEQ-package are in 
order at this point. The basic idea we employed was the following:
In the CTEQ package the parton distributions are given on a dynamical $x$- and 
$Q$-grid of variable size where the convolution of the kernels with the initial
distribution is performed on the $x$-grid. Due to the possibility of singular 
behaviour of the integrands, we perform the convolution integrals by first 
splitting up the region of integration according to the number of grid 
points, analytically integrating between two grid points $x_i$ and $x_{i+1}$ 
and then adding up the contributions from the small intervalls. We can do the 
integration analytically between two neighbouring gridpoints by approximating 
the distribution function through a second order polynomial $ay^2 + by +c$, 
using the fact that we know the function on the gridpoints $x_{i-1},x_i$ and 
$x_{i+1}$ and can thus compute the coefficients a,b,c of the polynomial. 
This approximation is warranted if the function is well behaved and the 
neighbouring gridpoints are close together. We treat the last integration 
between the points $x_1$ and $x_2$ (which are not to be confused with the 
$x_1$ and $x_2$ of the parton ladder) by taking the average of $x_1$ and 
$x_2$ and the values of the function at $x_1$ and $x_2$ and using those 
together with $x_1$, $x_2$ and the value of the function at $x_1$ and 
$x_2$ to compute the coefficients of the polynomial. The coefficients are
computed in the new subroutine NEWARRAY and the integration of the different 
terms in the kernels is performed in the new subroutine NINTEGR. The case 
$x_1 = \Delta = x$ is implemented analytically but separately in NINTEGR.
Appropriate changes in the subroutines NSRHSM, NSRHSP and SNRHS were made
to accomodate the fact that the kernels and also the integration routines 
changed from the original CTEQ package. A detailed description of the code
will be provided elsewhere \cite{AF97}.

\section{Limitations of the LLA in the nondiagonal case}
\label{sec:lim}

The LLA approach of the previous sections accounts for the
contribution of a certain rather limited range of integration in the 
parton distributions. Regions outside these limits might 
contribute to the leading power. Looking at some other physical
quantities such as $F_{2}$, where one finds substantial 
modifications due to the NLO-terms, we are forced to assume that this may
be also true in our case. This results in the urgent need to carry out a NLO 
calculation and numerical 
study of the evolution equation, which will be the next step of our program.

\section{Conclusions and Outlook}

In summary, we have calculated the evolution kernels for non-diagonal parton
distributions in the LLA using traditional methods and found agreement with
the results of \cite{Rad'96,Ji'96,B.B'88} deduced by other methods. 
It was important to show that the 
traditional approach can still be applied. Thus traditional methods can
be used to calculate systematically hard diffractive processes within 
the NLO approximations. We have also proved the similarity between the
diagonal and nondiagonal parton distributions. The latter ones determine
the cross sections of hard diffractive processes in the small $x$ region.
We have made predictions about the nondiagonal parton distributions
within the LLA with the help of a modified version of the 
CTEQ-package. Numerical calculations found the diagonal and 
nondiagonal gluon distributions, which dominate hard diffractive processes, 
to be very similar at small $x$ as expected from the previous discussion. 

\section{Acknowledgments}

We are indebted to A. Radyushkin for reading the paper, for many usefull 
comments and for pointing out some inconsistencies in an earlier version of 
this paper. This work was supported under grant number DE-FG02-93ER40771. 

\begin{figure}
\centering
\mbox{\epsfig{file=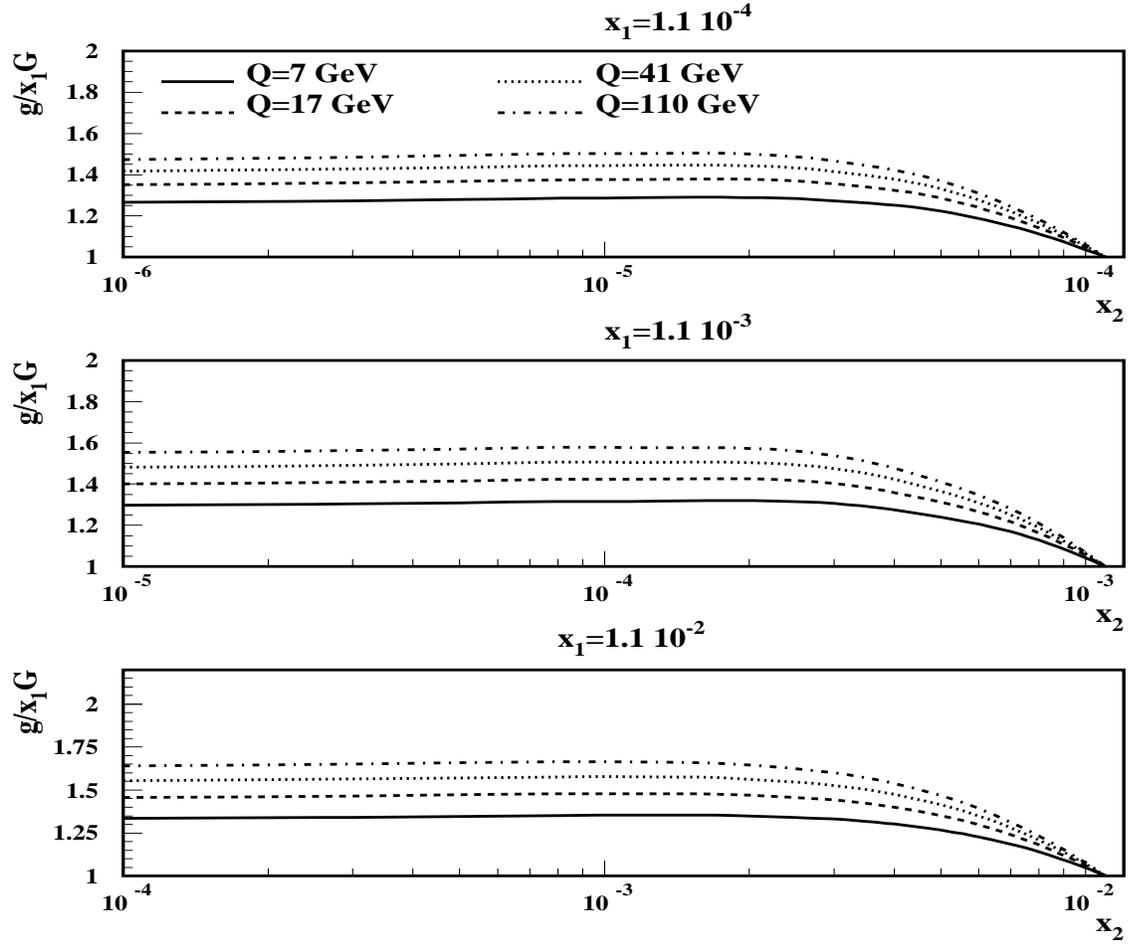,height=16cm,width=14cm}}
\caption{The fraction $g(x_1,x_2)/x_1G(x_1)$ as a function 
of $x_2$ for fixed $x_1$ and various energies $Q$. }
\label{Fig.1}
\end{figure}

\begin{figure}
\centering
\mbox{\epsfig{file=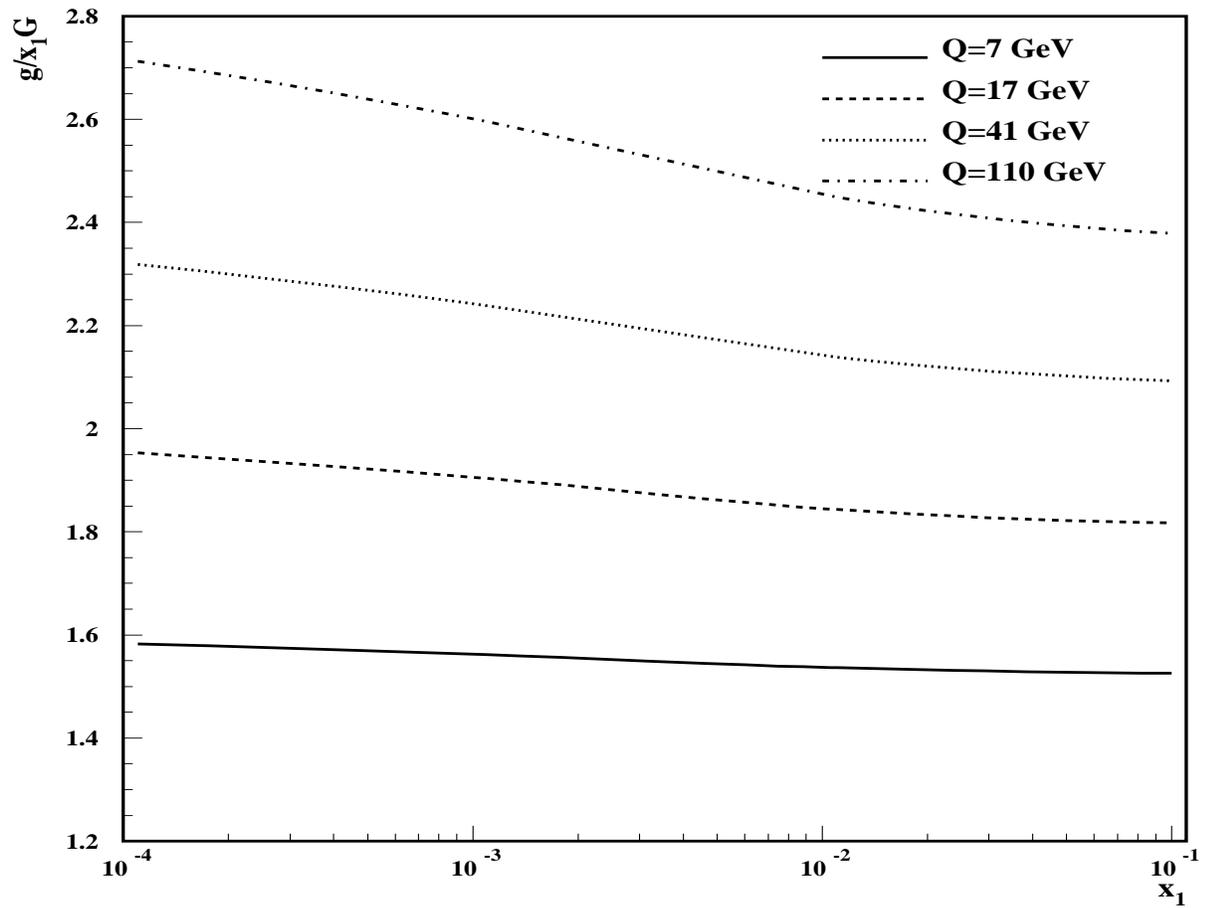,height=16cm,width=14cm}}
\caption{The fraction $g(x_1,x_2)/x_1G(x_1)$ as a function 
of $x_1$ for $x_2=0$ and various energies $Q$.}
\label{Fig.2}
\end{figure}

\end{document}